\documentclass[aps,pre,preprint,endfloats*,showpacs,groupedaddress]{revtex4}

\usepackage{graphicx}
\usepackage{amssymb}

\newcommand{\dtot}[2]{\frac{{\rm d} #1}{{\rm d} #2}} 
\newcommand{\dpar}[2]{\frac{\partial #1}{\partial #2}}  

\newcommand{\CLO}{\mbox{\rm ClO$_2^-$}} 
\newcommand{\TET}{\mbox{\rm S$_4$O$_6^{2-}$}} 
\newcommand{\HP}{\mbox{\rm H$^+$}} 
\newcommand{\SULF}{\mbox{\rm SO$_4^{2-}$}} 
\newcommand{\HSULF}{\mbox{\rm HSO$_4^-$}} 
\newcommand{\OH}{\mbox{\rm OH$^-$}} 
\newcommand{\EAU}{\mbox{\rm H$_2$O}} 

\begin{document}

\title{Dynamical Effects Induced by Long Range Activation in a Nonequilibrium
Reaction-Diffusion system}

\author{M. Fuentes}
\author{M.N. Kuperman}
\email[]{kuperman@cab.cnea.gov.ar}
\affiliation{Centro Atomico Bariloche and Instituto Balseiro,
8400 San Carlos de Bariloche, Argentina}
\author{J. Boissonade}
\email[]{boisson@crpp.u-bordeaux.fr}
\author{E. Dulos}
\email[]{dulos@crpp.u-bordeaux.fr}
\author{F. Gauffre}
\email[]{gauffre@crpp.u-bordeaux.fr}
\author{P. De Kepper}
\email[]{dekepper@crpp.u-bordeaux.fr}
\affiliation{Centre de Recherche Paul Pascal (CNRS), av. Schweitzer,
F-33600 Pessac, France}
\date{\today}

\begin{abstract}
We show both experimentally and numerically that the time
scales separation  introduced by long range activation can induce oscillations and
excitability in nonequilibrium reaction-diffusion systems that would
otherwise only exhibit bistability. Namely, we show that in the Chlorite-Tetrathionate
reaction, where the autocalytic species H$^+$ diffuses faster than the
substrates, the spatial bistability domain in the nonequilibrium phase
diagram is extended with oscillatory and excitability domains.  A simple
model and a more realistic model qualitatively account for the 
observed dynamical behavior. The latter model provides quantitative agreement with the
experiments.
\end{abstract}


\maketitle

Dynamics of spatial and spatio-temporal reaction-diffusion patterns in
chemical systems kept far from equilibrium by a permanent feed of
fresh reactants is an active domain of nonlinear science
\cite{FieldB,Shoka,Pojman}. These patterns include travelling waves in
oscillating or excitable media, Turing patterns, self-replicating
spots, labyrinthine patterns and dynamically structured fronts. They
result from the destabilisation of a  trivial state of high symmetry by a
positive feedback (`activation'), caused by autocatalysis or
substrate inhibition, competing with a negative feedback on
appropriate time scales. Although some patterns are
transiently observed  in some systems, these dissipative structures
can only be sustained in permanently fed open reactors.
Among these,
the One Side Fed Reactor (OSFR) made of a thin slab of a chemically
inert gel fed on one face and described further on in detail has
become popular. Recently, a new type of dynamical behavior, namely
spatial bistability, has been evidenced in this type of reactor
\cite{BispaD}. 

Up to now, two classes of spatial patterns have effectively been
observed.  In the first class, the primary source of the patterns is
the dynamical properties of the reaction in homogeneous
conditions. Reactions which oscillate or are excitable in homogeneous
conditions, i.e. in well stirred reactors, give rise to travelling
waves in unstirred media. The role of diffusion is to convert the
temporal dynamics into spatio-temporal dynamics \cite{note1}. The
phenomenon may occur whatever are the relative values of the diffusion
coefficients of the different species. In the second class, the source
of patterns is the so-called `long range inhibition'. An activatory
species, i.e. a species which controls the positive feedback, like an
autocatalytic species, diffuses much slower than the inhibitory
species. This favors the growth of the instability at localized
positions by creating two different time scales for spatial
interactions \cite{Mikhailov}.  The prototypes of such patterns are
the stationary Turing patterns \cite{Castets,TuringPhD,Ouyang}, but
long range inhibition is also involved in other phenomena like self
replicating spots \cite{self}, labyrianthine patterns \cite{EOE} or
morphological instabilities of fronts \cite{HV2,HV3,ArgBx1}. Since all
small molecules in solution have almost the same diffusion
coefficients, it was necessary to slow down the effective diffusion of
the activatory species by complexation with immobile sites to meet
these conditions experimentally \cite{Pearson,LengSc}.

Actually, `long range activation' where an activating species diffuses
faster than an inhibitory species defines a third class of
systems with several diffusive time scales.  Like with long range
inhibition, one expects this differential diffusion be responsible for
new instabilities that are not present in the sole chemical
kinetics. However, rather than inducing the growth of the
perturbations more or less locally, the fast diffusing activator tends
to excite perturbations at large distance from the initial
inhomogeneities and to transform a stationary state into
an excitable or an oscillatory one.  In a recent study of spatial
bistability in the chlorite-tetrathionate reaction, which exhibits
long range activation, we have reported a preliminary observation of
such a behavior \cite{Faraday}. Here, we report more extensive
experimental data and detailed numerical studies of these phenomena.

In a first part, we summarize former basic results on spatial
bistability in OSFRs, and present the reaction, the models and the
experimental systems. In a second part, we report the results obtained
with a simplified model to illustrate the general ideas.  In a third
part, we report both experimental results and  numerical
simulations with a more complete model. The last part is devoted to
the discussion of the results.

\section{Theoretical and experimental materials}
\subsection{Spatial bistability}
\begin{figure}
\includegraphics[width=8.5cm]{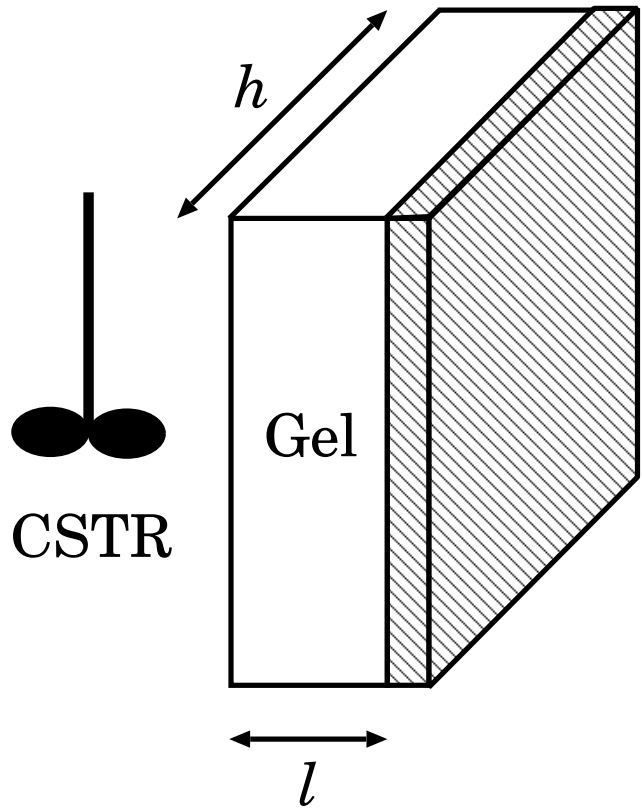}
\caption{Scheme of an OSFR.}
\label{OSFR}
\end{figure}

An OSFR is made of a film of a chemically resistant gel, of uniform
thickness $l$. One face is kept in contact with the
homogeneous contents of a continuous stirred tank reactor (CSTR). 
The other face is pressed against an impermeable wall (Fig.~\ref{OSFR}).  
The coupled
dynamical equations for the concentrations in the CSTR and in the gel
are respectively \cite{BispaD}:
\begin{equation}
\dpar{c_{i{\rm h}}}{t}\;=\; f_i({\bf c}_{\rm h})
\;+\;\frac{({c_{i0}}-{c_{i{\rm h}}})}{\tau}
\;+\;\rho_V\frac{D_i}{l}\left(\dpar{c_i}{r}\right)_{r=0} \label{eqcstr}
\end{equation}
and 
\begin{equation}
\dpar{c_i}{t} \; =\; f_i({\bf c}) \; +\;D_i\nabla^2c_i \label{eqgel}
\end{equation}
where $c_{i0}$, $c_{i{\rm h}}$ and $c_i$ are the concentrations of
species $i$ respectively in the input flow, in the CSTR, and inside
the gel, $D_i$ is the corresponding diffusion coefficient, $\tau$ the
residence time of the reactor , $\rho_V$ the ratio of the volume of
the gel to the volume of the CSTR, and $r$ the distance to the
CSTR/gel interface.  The $f_i$'s are the reaction rates.  In the right
hand side of Eq.~(\ref{eqcstr}), the three terms represent the change
of the concentrations per time unit.  The first term gives the
contribution of the reaction. The second one represents the input and
output flows of the species. It contains all the expandable control
parameters of the system, namely $\tau$ and the $c_{i0}$. The third
term results from the diffusive flux of the species at the interface
between the gel and the CSTR and represents the feedback of the gel
contents on the CSTR dynamics. When the volume of the CSTR is large in
regard to the volume of the gel ($\rho_V<<1$), this last term can
generally be dropped so that the chemical state of the CSTR is
independent of the state of the gel and the concentrations in the CSTR
define a Dirichlet boundary condition for Eq.~(\ref{eqgel}) at $r=0$,
whereas a no-flux boundary condition is applied at $r=l$. To avoid
ambiguities, let us make more precise the terms we use. Whereas in
usual OSFRs the thickness $l$ is small in regard to the two other
dimensions, in this paper, we want to investigate the behavior of the
system along the direction orthogonal to the faces in contact with the
CSTR and the impermeable wall.  Thus, we shall consider systems made of
`flat' gels in which one of the dimensions parallel to these faces ($h$
in Fig.~\ref{OSFR}) is still smaller and negligible. Such systems can
be considered as two-dimensional.  Nevertheless, the size $l$, in the
direction orthogonal to the faces, 
remains small in regard to the second direction
parallel to these faces. Size $l$ and the direction orthogonal to the
faces will be further referred to as the `depth'
or the `width' of the gel, the negligible dimension $h$ becoming
the `thickness'.

A number of reactions which present autocatalytic steps are bistable
in a CSTR \cite{FieldB,Pojman,GrayScott}. 
These reactions normally exhibit `clock' behavior in batch conditions,
i.e. a more or less long induction period followed by a fast switch to
the equilibrium. In these reactions, the transformation rate of
the substrate(s) tremendously  increases during the switch which
occurs when the level of the autocatalytic species is large enough.
For residence times much
shorter than the typical reaction time, i.e. at high flow rates, the
composition inside the CSTR is close to the composition in the input
flow. It is referred as the `Flow state' or state F.  For residence times
much larger than the typical reaction time, i.e. at low flow rates,
the reaction is almost completed in the CSTR and the composition in
the CSTR is close to a thermodynamic equilibrium. This is referred as
the `Thermodynamic state' or state T. For intermediate flow rates, both
states are stable for the same composition of the input flows.  Spatial bistability can be observed
when such a reaction is operated in an OSFR, the CSTR of which is set
in the flow state.  Although this phenomenon, which extends the concept
of bistability to space, has been previously described in simple models
\cite{GrayScott} and in real experiments \cite{CouetBx,Couetall}, its
importance has been recognized only recently and a detailed analysis
was worked out in relation with the development of the OSFRs
\cite{Bispa,BispaD,BispaT}. 

To understand the basic principles of spatial bistability, let us
consider the concentration profile of a substrate of concentration
$X$ and diffusion coefficient $D$ consumed in the reaction and let
us imagine that the depth $l$ of the gel can be changed in a
continuous way.  The time taken by fresh reactants fed by diffusion at the
CSTR boundary to reach the opposite boundary is of the order of $t_D =
l^2/D$.  We start from $l=0$ and increase $l$ adiabatically so that
the system is always in its asymptotic stationary state.  Let $t_i$ be
the induction time for the CSTR composition (in state F).  While $l$
is small enough for $t_D \ll t_i$ to hold, the whole volume of the gel
remains in state F. We shall still refer this state of the gel as a F
state. When $t_D$ reaches a value of the order of $t_i$, for a value
$l=l_{\rm max}$, the fast reaction starts at the impermeable wall and
a steep front connecting the F state to a T state forms. Such a state
will be referred as a `mixed state' or FT state. But, the
amount of the autocatalytic species is now large at the impermeable wall of
the gel ($r=l$) and can diffuse backwards switching on the reaction,
so that the front can recede to a value $l=\delta<l_{\rm max}$.
Actually, if one assumes that the reaction rate is
extremely large inside the front, one can neglect the rate of
the consumption of the substrate elsewhere and the concentration
of the latter can be approximated by the solution of a diffusion equation 
between a fixed value at $r=0$ (the CSTR) and $r=\delta$ (the front).
Then, it can easily be shown that:
\begin{equation}
\delta \approx \frac{D_i|\Delta X|S}{\dot{Q}} \label{valdelta}
\end{equation}
where $\dot{Q}$ is the total amount of substrate converted by time
unit in the gel and $\Delta X$ is the variation of concentration
between the F and the T states \cite{BispaD}. This value is
practically independent of $l$, so that one can now decrease $l$ down
to a value $l_{\rm min} \sim \delta $ before the system switches back
from the FT to the F state. Thus, for $l_{\rm min} < l < l_{\rm max}$,
both concentrations profiles F and FT are stable. This defines the
spatial bistability. The extent of the bistability domain depends on
the concentrations in the CSTR -- always close to those in the input
flow -- and the residence time, so that various nonequilibrium phase
diagrams can be built \cite{BispaD, BispaT, Faraday}. In spite of its
apparent crudeness, this theory gives accurate results.
Extensive studies have been performed experimentally and numerically
on the Chlorine Dioxide-Iodide  (CDI) reaction.
Up to now, differences in diffusion coefficients did not seem
to play any role. Actually, whether the system is ruled by equal diffusion
coefficients or by long range inhibition, it exhibits a standard
spatial bistability. We shall see that the situation is fairly more
complex when long range activation comes into play.

\subsection{The Chlorite-Tetrathionate reaction}
The  Chlorite-Tetrathionate (C-T) reaction  has
essentially been studied in closed systems for the determination of
stoichiometric and kinetic aspects \cite{Nagypal,HV1} as well as a
source for propagating cellular fronts \cite{HV2,HV3,ArgBx1}.  In
excess of chlorite, the reaction kinetics is well approximated by the
following overall balance equation
\begin{equation}
\mbox{ 7 ClO$_2^-$ + 2 S$_4$O$_6^{2-}$ + 6 H$_2$O}
\longrightarrow \mbox{7 Cl$^-$ + 8 SO$_4^{2-}$ + 12 H$^+$}\label{kin}
\end{equation}
and the reaction rate by
\begin{equation}
v_R=-\frac{1}{7}\dtot{\mbox{[ClO$_2^-$]}}{t}
=k\mbox{[ClO$_2^-$][S$_4$O$_6^{2-}$][H$^+$]$^2$} \label{kinlaw}
\end{equation}
Since, during the reaction, large pH changes are observed, the fast
dissociation equilibria of $\EAU$ and $\HSULF$ have also to be
accounted for. This rate law exhibits quadratic autocatalysis in
[$\HP$]. Although the validity of this rate law has been determined
for a limited range of pH (4.7-5.5) \cite{Nagypal}, it has been
successfully used in much larger domains. Nevertheless, the original
value $k=10^9$~M$^{-3}$s$^{-1}$ had to be adjusted to fit with
experiments.  The value $k=5\times10^6$~M$^{-3}$s$^{-1}$ has recently
been proposed to fit the experimental data of the bistability domain
of the reaction in a CSTR \cite{Faraday} and will be used in the
following.

It is well known that, whereas most species have a diffusion
coefficient close to 10$^{-5}$~cm$^2$s$^{-1}$, the autodiffusion
coefficient of the proton H$^+$ is almost ten times
larger. Nevertheless, accounting for the electroneutrality condition
and the presence of the slowest ions, the effective diffusion of H$^+$
is  lower  \cite{Cussler}. Solving
reaction-diffusion equations that include all ionic effects is a
formidable task which has been only performed on simple models
\cite{Marek}. For practical reasons, it is reasonable to use a
constant and uniform effective value for each diffusion coefficient.
We shall assume that all diffusion coefficients are equal to a unique
effective value $D_1$, except for H$^+$ and OH$^-$ which have both a
higher autodiffusion coefficient and are closely related. Let $D_2$ be
a common value for these two species.  With these
assumptions, it was shown in ref \cite{Faraday} that the experimental
data can be correctly interpreted only if $d=D_2/D_1$ ranges slightly
above 3. Thus it is necessary to take into account that the autocatalytic
species H$^+$ diffuses much faster than the inhibitory species, so
that the reaction-diffusion system enters the class of long range
activation.
 
\subsection{The simple model}
In order to make extensive computations, we shall first define a simple model,
the properties of which are close enough to the original system to
draw significant conclusions. Our main purpose is that, due to its 
formal simplicity,
this model can be further used as a prototype for long range activation.
The model relies on two strong assumptions:
a) The kinetics is completely described by Eqs.~(\ref{kin}, \ref{kinlaw}).
b) The concentrations at the CSTR boundary are fixed to their values in
the input flow, i.e. we assume a residence time $\tau=0$ (infinite flow rate).
In these conditions, the reaction-diffusion system is:  
\begin{eqnarray}
\dpar{\CLO}{t} & = & -7k[\CLO][\TET][\HP]^2 +D_1\nabla^2[\CLO] \nonumber \\
\dpar{\TET}{t} & = & -2k[\CLO][\TET][\HP]^2 +D_1\nabla^2[\TET] \label{model1}\\
\dpar{\HP}{t} & = & 12k[\CLO][\TET][\HP]^2 +D_2\nabla^2[\HP] \nonumber
\end{eqnarray}
In this paper, we do not consider short transients that follow a change 
of parameters and only consider the asymptotic regime. Eliminating the 
reaction term between the two first equations, one is
left with a diffusion equation with a fixed boundary condition at
$r=0$. In the long time limit, the
solution is $2[\CLO]-7[\TET]=2[\CLO]_0-7[\TET]_0$, where index 0 holds
for the concentrations fixed at the boundary. Then [\TET] can be expressed
as a function of [\CLO] and can be
eliminated from Eqs.~(\ref{model1}). One gets the two-variable reduced system:
\begin{eqnarray}
\dpar{u}{t^\prime} & = & -(u-\xi)uv^2 +\nabla^2_{r^\prime}u \nonumber
\\ & &\label{toy}\\ \dpar{v}{t^\prime} & = & \frac{12}{7}(u-\xi)uv^2
+d\nabla^2_{r^\prime}v \nonumber\end{eqnarray} where we have
introduced the new time and space scales $t^\prime=2k([\CLO]_0)^3t$,
$r^\prime= (2k/D_1)^{1/2}([\CLO]_0)^{3/2}r$ and the following
normalized variables and parameters: $u=[\CLO]/[\CLO]_0$,
$v=[\HP]/[\CLO]_0$, $d=D_2/D_1$ and
$\xi=1-(7/2)[\TET]_0/[\CLO]_0$. This system is formerly identical to
the system used in  previous studies in closed systems \cite{HV1}. 
According to the
definition of $u$, one has $u=1$ at the CSTR boundary. Thus, the
control parameters are $v_0$, $d$, and $\xi$, where $v_0$ is the fixed
value of $v$ at the CSTR boundary.  Parameter $\xi$ represents the
normalized excess of chlorite. Provided it is not too large -- this is
the case in most experiments -- it was found to play a minor role in
the model and does not bring qualitative changes in the results. Thus,
it was fixed at $\xi=0$ which would correspond to exact
stoichiometric proportions of chorite and tetrathionate.

Note that, if $d=1$, the third equation could be eliminated in the
same way and the system would be reduced to a unique variable $u$. A
similar reduction to the same one-variable equation could also be done for a
homogeneous CSTR.  A one variable system can be bistable but
never oscillating or excitable, since two dynamical time scales at least are
necessary. This suggests that the C-T reaction can be neither
oscillating or excitable in a CSTR. More sophisticated models support
this assertion, which is confirmed by experiments. 

This model suffers from serious limitations. First, at the CSTR
boundary, we use fixed concentrations equal to the concentrations in
the input flow.  Since, in the flow state, the concentrations should
be close to those in the input flows, this seems to be a reasonable
approximation. Nevertheless, at finite residence times, the values of
$v_0$ could correspond to non existent or unstable states of the CSTR.
The approximation is strictly valid only in the practically
unreachable limit of infinite flow rates ($\tau=0$). In the same way,
one can fix $v_0$ to arbitrary high values for which the CSTR would
necessarily be in a T state. This can lead to ambiguities in the
definition of the spatial states F and T and to experimentally
unreachable regions in  phase diagrams so that some care is
necessary for their interpretations.  Finally, values of $v_0$ can also
be taken arbitrary low, down to zero, an unrealistic situation since
the dissociation equilibrium of water imposes that $[\HP]$ remains
positive and finite. Thus, this model must be only thought as a `toy
model', with the advantage that the kinetics remains closely related
to that of a real system.  This will allow us to evidence dynamical
phenomena induced by long range activation. Such models have also
been recently used by Benyaich et al. in studies of spatial bistability
\cite{Khalid}.
To make direct comparison
with experiments, a more elaborated model has to be devised.

\subsection{The extended model}
To avoid oddities due to the simple model oversimplifications, one has
to  account, not only for the dynamics of the CSTR at $r=0$
included in the Eq.~(\ref{eqcstr}), but also for the way in which the H$^+$
ions are introduced in the reactor and the equilibria in which they
are involved.  In the extended model that we used to make comparisons
with the experimental results, the following fast equilibria are taken
into account in addition to Eq.~(\ref{kin}):
\begin{equation}
\HP+\OH \rightleftharpoons \EAU \label{kineau}
\end{equation}
with rate law
\begin{equation}
v_e=k_e^+-k_e^-\;[\HP][\OH]
\end{equation}
and
\begin{equation}
\HP+\SULF \rightleftharpoons \HSULF \label{kinsulf}
\end{equation}
with rate law
\begin{equation}
v_a=k_a^+\;[\HSULF]-k_a^-\;[\HP][\SULF]
\end{equation}
Discarding the species that do not contribute to kinetics in
Eqs.~(\ref{kin}, \ref{kineau}, \ref{kinsulf}), one retains the six
variables $[\HP]$, $[\OH]$, $[\CLO]$, $[\TET]$, $[\SULF]$ and
$[\HSULF]$.  The following values are used for the kinetic constants:
$k_e^-=1.4\times 10^{11}$ M~s$^{-1}$, $k_e^+=K_e\;k_e^-$ with
$K_e=10^{-14}$, $k_a^-= 10^{11}$ M~s$^{-1}$, $k_a^+=K_a\;k_a^-$ with
pK$_a=-\log (K_a)=1.94$.  The diffusion coefficient $D_1$ is fixed
to a standard value $D_1$=10$^{-5}$ cm$^2$s$^{-1}$ and the fast
species diffusion coefficient $D_2$ is fixed to $D_2=3.4\times
10^{-5}$ cm$^2$s$^{-1}$ to fit at best the experimental data. Accurate
one-dimensional numerical solutions of the resulting
reaction-diffusion system have been obtained by an implicit finite
difference method for stiff systems with adaptative grid in time and
space. Unfortunately, due to the higher complexity and the extreme
stiffness of the model, it was impossible to make systematic studies
of this model in higher dimensions. Two-dimensional computations were
limited to a few points in parameter space.  In order to avoid
problems in the definition of the input concentrations and to make
easier the comparisons with the experimental data and a previous work,
we use, as the main control parameter, the quantity $\alpha$ 
defined in the  experimental section below.

\subsection{Experimental conditions}
In order to resolve the concentration profiles in the depth $l$ of the
reactor and to minimize both technical difficulties and parasitic edge
effects, we have developed one-side-fed flat annular reactors,
i.e. annular OSFRs.  Complete descriptions and technical details on
such reactors can be found in ref.~\cite{BispaD}. These OSFRs are made
of a flat annular piece of 2\% agarose gel, of thickness $h \leq$0.5
mm. They have a fixed outer radius $R_{\rm out}$=12.5~mm and an inner
radius $R_{\rm in}$ such that the width $l=R_{\rm out}-R_{\rm in}$
ranges from 0.5 to 3~mm. The agarose gel can be considered as
chemically inert in regard to the reaction. In this soft hydrogel, the
diffusion coefficients are close to those in pure water. 
The ratio $R_{\rm in}/l$ is large enough for the curvature
effects to be negligible. The inner rim of the annulus is tighten against
the impermeable boundary while the outer rim is in contact with the
contents of a CSTR of 25~cm$^{3}$, fed through a single inlet port
with appropriately premixed solutions of reagents stored in separated
reservoirs.  The residence time of the reactor $\tau$=600~s, the
temperature T=25$^0$C, and the input flow concentrations of sodium
chlorite [NaClO$_2$]$_0=1.9\times 10^{-2}$~M and potassium
tetrathionate [K$_2$S$_4$O$_6$]$_0$=0.5$\times 10^{-2}$~M are kept
fixed during all the experiments.  Besides the two above mentioned
reagents, variable mixtures of perchloric acid and sodium hydroxide
are added to the input flow of reagents to control the pH of the
premixed feed solution. This is done by pumping different volume
ratios of acid and base, keeping the overall flow rate constant. For
experimental and graphic convenience, this relative distribution is
characterized by a control parameter $\alpha$ on which the acid and
base flow concentrations depend linearly: one has [HClO$_4$]$_0$ =
$\alpha\times0.67~10^{-2}$~M and [NaOH]$_0$ =$(1-\alpha)\times
3.33~10^{-2}$~M.  The pH of the feed mixture decreases from a basic to
an acid state as $\alpha$ increases from 0 to 1. The parameter
$\alpha$ effectively controls the pH of the solution and is the only
expandable control parameter in this series of experiments.  The
patterns are made visible by bromophenol blue (1\% solution), a pH
sensitive indicator that turns from dark purple in the basic state to
pale yellow in the acid state.  Patterns are monitored by a CCD
camera connected to a time lapse video recorder.
 
The dynamics of the C-T reaction operated in
the CSTR exhibits bistability in the range $0.18<\alpha<0.84$
between a F branch (with pH values typically above 9) and a T branch (with pH
values typically below 3). The upper limit corresponds to the value of
$\alpha$ where the input flow switches from a basic to an acid solution.
The lower limit was used to adjust the rate constant $k$ to
the value $k=5\times10^6$~M$^{-3}$s$^{-1}$ \cite{Faraday}. 

\section{The simple model analysis}

In the preliminary experiments of ref.~\cite{Faraday},
we had observed, in addition to spatial bistability,
oscillations of the FT state front and excitability
phenomena in the domain where the sole F state is stable.
While the stability of concentration profiles can
be determined with 1-D simulations, the latter
needs 2-D computations. Thus, the stability limits are
determined in 1-D calculations by the same accurate implicit method as
the full model and the existence of excitability domain is studied in
two dimensional simulations, performed
with two different numerical methods, namely an implicit
finite difference hopscotch method with automatic step
control and a finite element method.

\begin{figure}
\includegraphics[angle=-90,width=8.5cm]{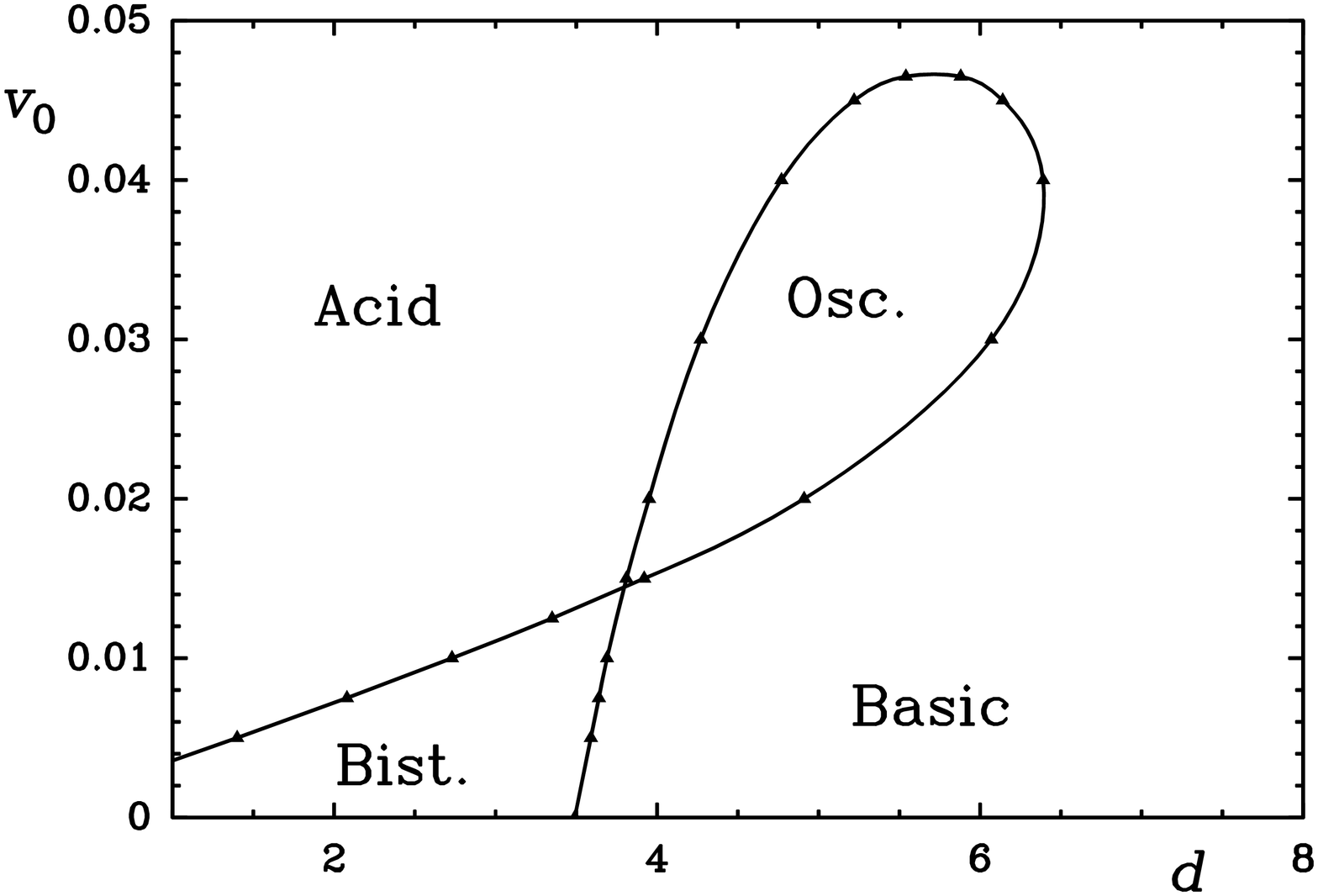}
\caption{Non equilibrium phase diagram of the simple model at $l$=10 in the 
plane $(d,\;v_0)$.}
\label{diagdv}
\end{figure}

\begin{figure}
\includegraphics[angle=-90,width=8.5cm]{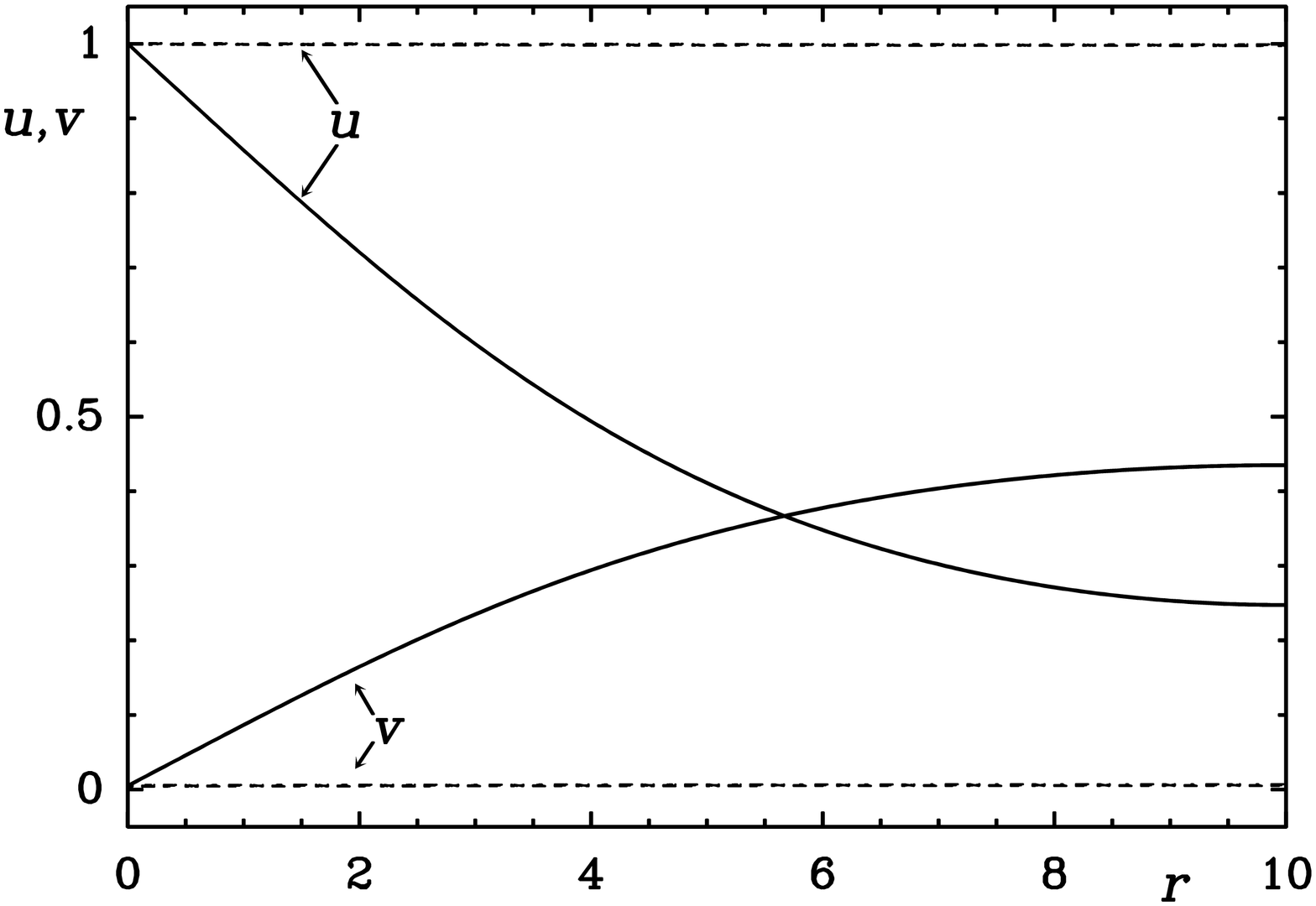}
\caption{Spatial bistability in the simple model: profiles of $u$ and
$v$ as a function of depth.\\Dotted lines: F state ; full lines: FT
state. Parameters: $l$=10, $d$=3, $v_0$=0.005.}
\label{proftoy}
\end{figure}

First, choosing a typical depth value $l=10$ (expressed
in reduced units precedently defined), we have built a
stability diagram in the plane $(v_0,d>1)$ (Fig.~\ref{diagdv}).
At low
values of $v_0$ and $d$, there is a large region of spatial
bistability that extends down to $v_0=0$. When $v_0\longrightarrow 0$
the `acid' state must obviously be obtained by a continuous
decrease of $v_0$ from another `acid' state or by an acid perturbation of
the system. 
In Fig.~\ref{proftoy}, 
we have represented the two stable
concentration profiles through the depth of the gel for a given set of
control parameters. Contrary to the standard case, in the FT state,
these profiles are always monotonous and do not exhibit a steep
front. The states are nevertheless well characterized by the
low (`basic') or high (`acid') value of $v$ at $r=l$,
the impermeable boundary of the gel. When $d \gg 1$, the system
exhibits temporal oscillations in place of bistability. Since the
system cannot oscillate in homogeneous conditions, these oscillations
clearly result from the different time scales introduced by the
differential diffusion. Although the problem is of spatial nature, the
stability limits present the characteristic topology of the so-called
`cross-shaped diagram' that connects bistability and oscillations of
numerous reactions in homogeneous conditions \cite{crossshape}.

\begin{figure*}
\includegraphics[width=14cm]{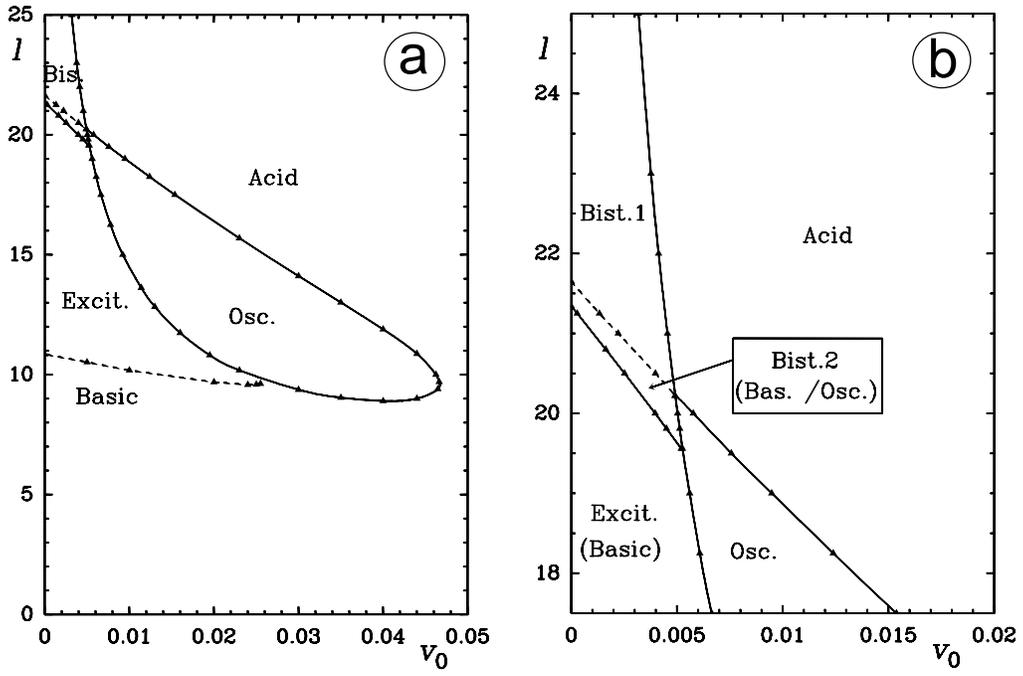}
\caption{Non equilibrium phase diagram of the simple model at $d=5.5$
in the plane $(v_0,\;l)$. a)~Large view b)~Detail of a.}
\label{diagvl}
\end{figure*}
\begin{figure}
\includegraphics[width=8.5cm]{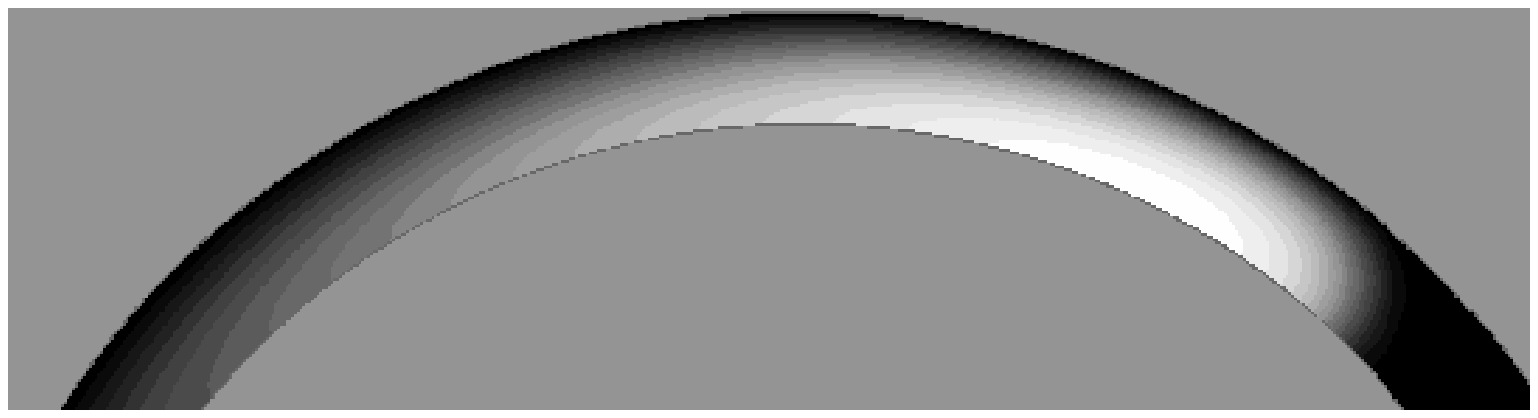}
\caption{Excitability in the simple model: travelling pulse.
$d=5.5$, $l=17, v_0=0.003$. Concentration map of $v$: from minimum (black)
to maximum (white).}
\label{pulse}
\end{figure}

A second diagram in the plane $(v_0,l)$ has been extensively computed
for $d$=5.5, a value chosen for the richness of the dynamical
situations exhibited at this diffusion ratio (Fig.~\ref{diagvl}).  For
small depths, the system is always in the flow (or `basic') state, as
expected. At intermediate values of depth, there is a large domain of
oscillations. It vanishes at large $v_0$ values for which one
continuously changes from the `basic' to the `acid' state when $l$ is
increased. For large depths, the FT (or `acid') state is again present
down to $v_0=0$, which defines a domain of spatial bistability at
small $v_0$ values. The most interesting part of this diagram is the
region labeled `Bist.2' in Fig.~\ref{diagvl}(b).  In this narrow
region, the system still exhibits spatial bistability, but the `acid'
state becomes oscillating. Thus, the bistability domain is divided in
two regions, one of standard bistability between two stationary
concentration profiles (`Bist.1'), and a second one between a
stationary profile and an oscillating one (`Bist.2'). Moreover, when
one decreases $l$ and crosses the stability limit of the oscillatory
`acid' state, the unique stable state, namely the F (`basic') state,
exhibits particular properties. If, in this stable state, an acid
perturbation (such as $v=1$) is applied close to the impermeable wall,
the perturbation does not smoothly decay to the stable state but
stimulates a further increase of $\HP$ ions concentration in its
vicinity. A pair of pulses forms and propagates along the wall in
opposite directions.  After relaxation of transients, they travel with
constant shape, amplitude and velocity, which precisely defines the
phenomenom of excitability.  The limits of this excitability domain
are reported in Fig.~\ref{diagvl}(a). This domain extends all the way
to the oscillation zone.  In the vicinity of this latter zone, the
recovery tail of the pulse can become oscillatory in space. Although,
for more efficiency, these limits have been accurately computed in a
rectangular reactor, we have also also performed some runs in an
annulus, analog to that of the experimental device. An example of an
excitation pulse propagating in such an annulus (partial view) is
represented in Fig.~\ref{pulse}.
The limits of this excitability domain are
reported in Fig.\ref{diagvl}(a). This domain extends to the
oscillation zone.  In the vicinity of this latter zone, the recovery tail of
the pulse can become oscillatory in space.

\section{Extended model analysis and experimental results}
In this section, we report the numerical results that have been obtained
with the extended model and simultaneously make a direct comparison
with the experimental data. We consider the existence and stability of
the F state and the FT state in the plane ($\alpha,l$).
 
\begin{figure}
\includegraphics[width=8.5cm]{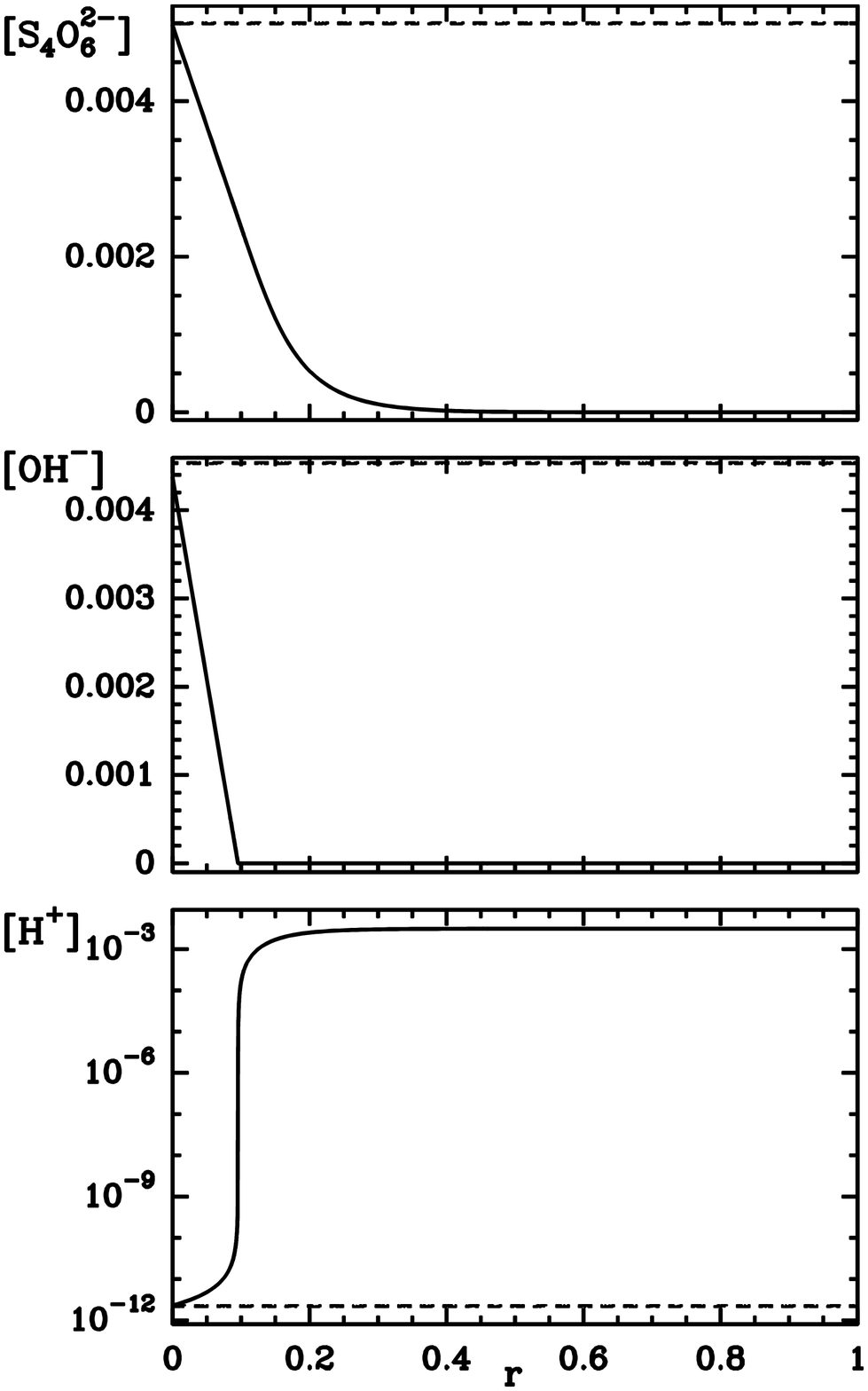}
\caption{Computed concentration profiles through the gel depth.
Parameters: $l$=1mm, $\alpha$=0.72. Dotted lines: F state; Full lines:
FT state.}
\label{numprof}
\end{figure}

Spatial bistability is observed over a significant domain of the plane.
For instance, in Fig.~\ref{numprof}, we show the two stable
concentration profiles (F and FT states) computed for a same set of
control parameter corresponding to a point located inside the bistability
domain. 
Contrary to the case of the simple model, the FT state exhibits
a steep front.
The concentration profile of the autocatalytic species $\HP$
is plotted in logarithmic scale to emphasize on the position and the
stiffness of the front. The concentration of $\TET$ and OH$^-$ 
are represented in linear coordinates to underline the linear drop
of the substrates concentrations as predicted by
Eq.~(\ref{valdelta}).

In experiments, the pH  color indicator reveals the two different states
\cite{Faraday}. In the FT state,
the gel is found to remain in a basic state only in a narrow band located along
the outer rim. The nature of the different dynamical states were determined
unambiguously in the whole parameter range that was explored.

\begin{figure}
\includegraphics[width=8.5cm]{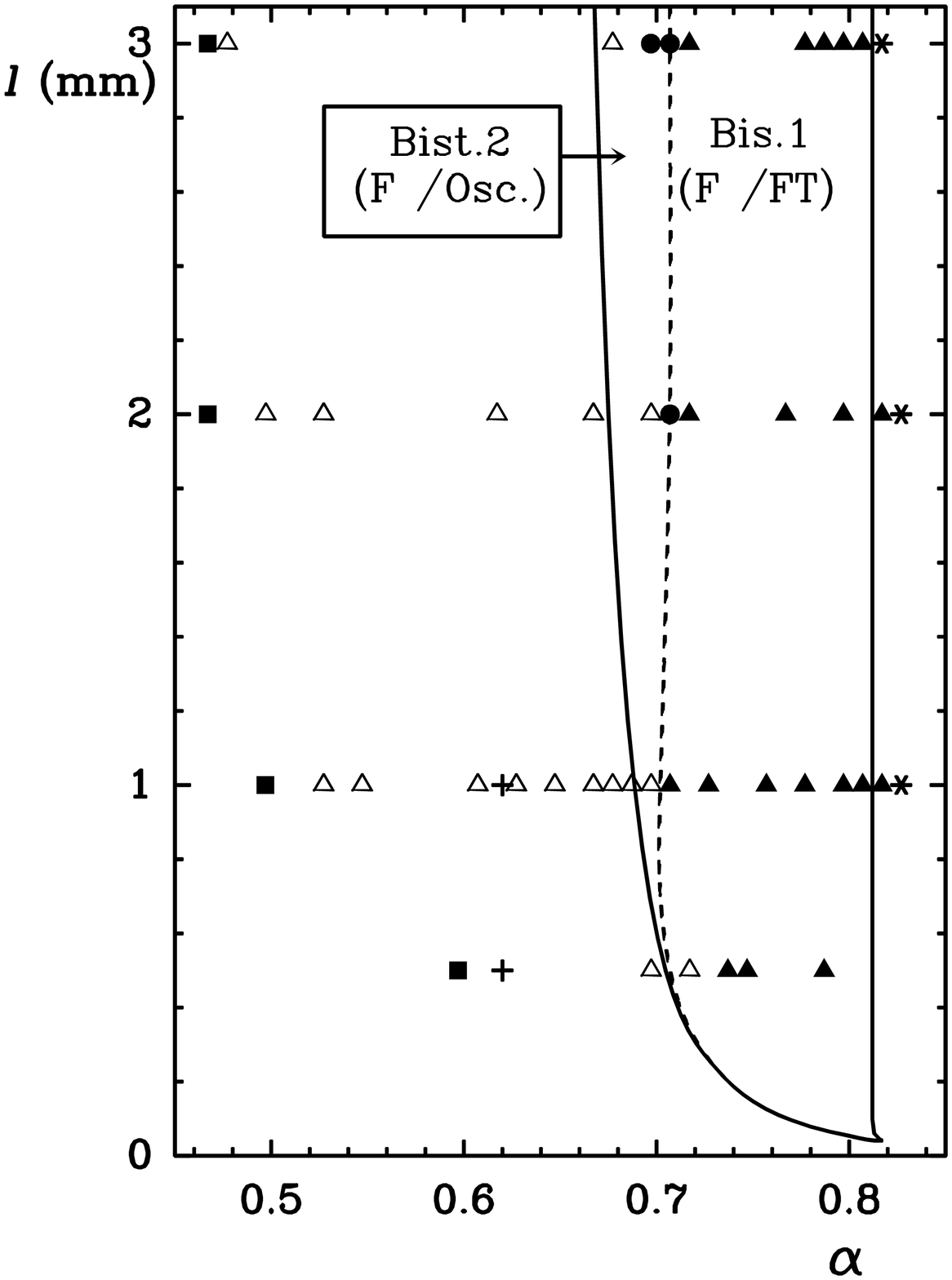}
\caption{Non equilibrium phase diagram in the plane ($\alpha$, $l$).
Experimental points: $\blacksquare$ monostable with non excitable F
state, $\triangle$ monostable with excitable F state, $\bullet$
bistable with oscillatory FT state, $\blacktriangle$ bistable with
stationary FT state, $\ast$ the CSTR switches to the thermodynamic
state. Computed results: the full lines are the limits of the
bistability domain, the dotted line is the limit of the oscillory FT
state within the bistability region, + marks are points for which the
F state has been shown to be excitable.}
\label{diag34}
\end{figure}

In Fig.~\ref{diag34}, we report the computed limits of the
bistability domain (full lines) in the plane ($\alpha,l$) and the
experimental points. 
Gel annuli with $l<0.5$ mm could not be technically achieved, so that
this part of the diagram cannot be explored
experimentally. Numerically, the upper limit $\alpha_{\rm max}$ of the
FT state is slightly lower than the limit $\alpha_{\rm T}=0.833 $
computed for the F state in the CSTR.  Actually, in the FT state, the
concentration gradients of the substrates at $r=0$ are large as seen
in Fig.~\ref{numprof}, so that, although $\rho_V \ll 1$, the last term
in Eq.~(\ref{eqcstr}) is not completely negligible and the contents of
the CSTR are entrained into the T state slightly before the limit
$\alpha=\alpha_{\rm T}$ is reached. Experimentally, this minor deviation
is too small to be detected.  There is a first major difference with
the standard CDI reaction \cite{BispaD}: The spatial bistability
persists at large values of $l$, which seems in contradiction with our
theoretical presentation asserting that the F state cannot exist when
the width of the gel is large. Actually, this results from the extreme
slowness of the reaction when the system is in the basic F state.
This leads to a quasi infinite induction time. Thus, one
cannot switch spontaneously from the F state to the FT state by a
continuous change of $l$, if one remains within a realistic range.
Nevertheless, one reaches this state by an appropriate perturbation or
choice of the initial condition in the gel, but the bistability
domain extends to huge values of the depth $l$.

\begin{figure}
\includegraphics[width=8.5cm]{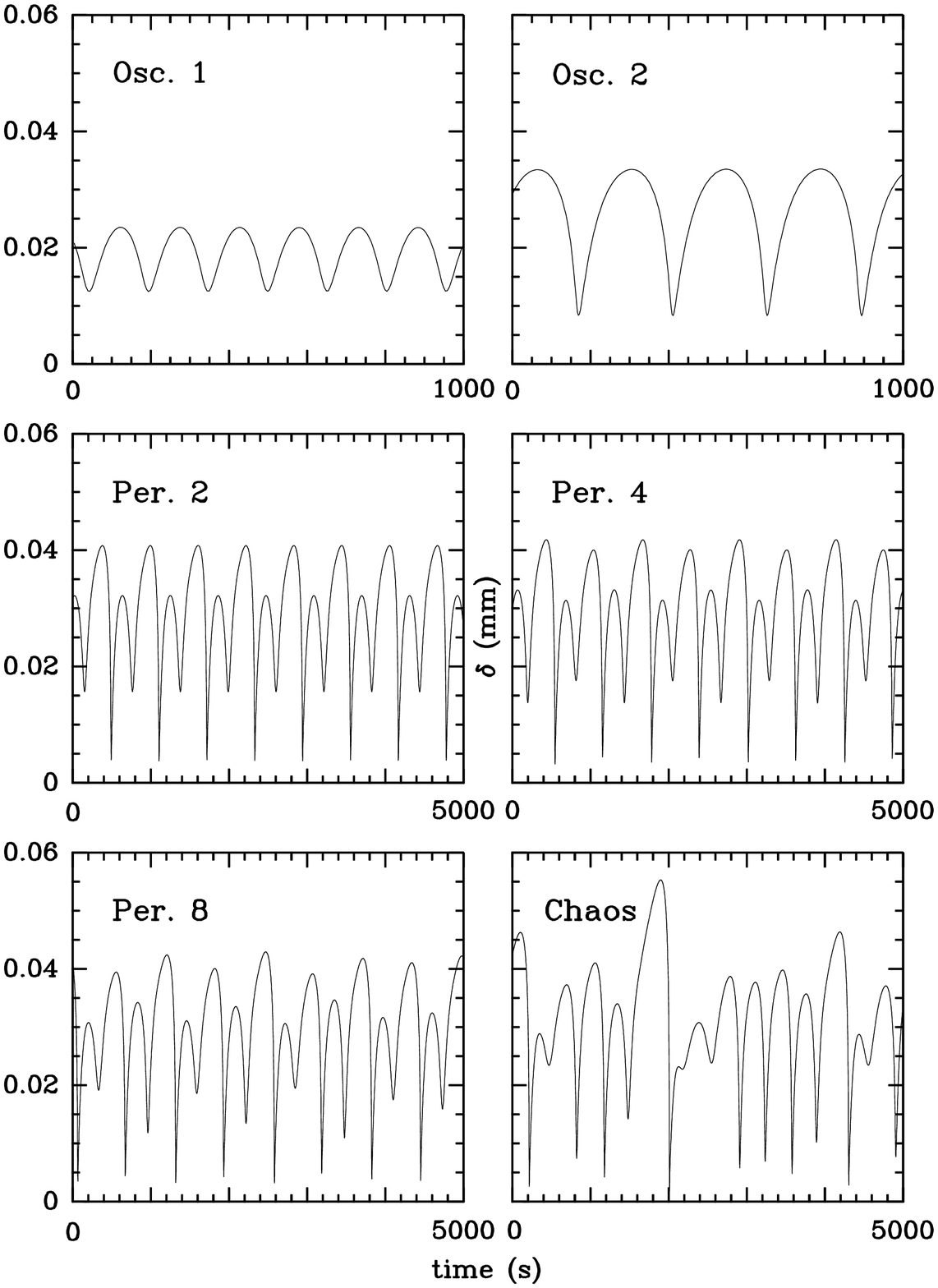}
\caption{Computed oscillations of the front position $\delta$ ($l$=1mm):
Oscillation 1: $\alpha$=0.700 ; Oscillation 2: $\alpha$=0.692;
Period 2: $\alpha$=0.6895 ; Period 4: $\alpha$=0.6894;
Period 8: $\alpha$=0.6892 ; Chaos: $\alpha$=0.6885.}
\label{oscil34}
\end{figure}

Contrary to the standard case of the CDI reaction, the limit of the FT
state when $\alpha$ decreases at constant $l$ is complex and exhibits
dynamical phenomena both numerically and experimentally, provided that
$l$ is large enough as seen in Fig.~\ref{diag34}. When the value of
$\alpha$ decreases and crosses the dotted line, the position of the
front begins to oscillate and $\delta$ becomes a function of time.
This domain of oscillatory dynamics is narrow and difficult to explore
precisely in experiments.  However, one dimensional numerical
simulations allow for a precise exploration of this oscillatory
domain.  In Fig.~\ref{oscil34}, we represent the numerical
oscillations for decreasing values of $\alpha$ at $l$=1~mm, but
similar sequences are observed for other widths of the gel.
When
$\alpha$ decreases, the amplitude of the oscillations increases and
they take a more and more relaxation character.  When the $\alpha$ values
are drawn still closer to the
stability of the FT state, one observes, over a very narrow
range of $\alpha$ values, a period doubling sequence leading to chaotic
oscillations. When $\alpha$ decreases from the onset of oscillations,
the period of the oscillations slowly increases. On the contrary, the period
strongly increases when the width $l$ increases.  Since the
experimental conditions are never strictly uniform all along the
outer rim of the gel and since the width $l$ cannot be perfectly
constant, the oscillations of the front lead to the formation of
pacemakers that emit travelling waves with periods comparable to those
found in the computations.

\begin{figure}
\includegraphics[width=8cm]{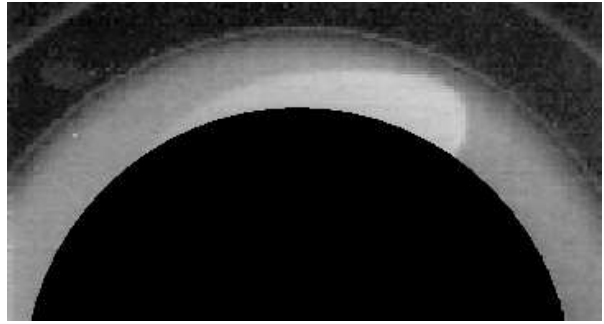}
\caption{Excitation pulse (experimental): an acid pulse (clear color)
propagates to the right along the
inner rim. $l$=0.3mm, $\alpha$=0.48.}
\label{excit}
\end{figure}

Beyond the stability limit of the FT front, dynamical phenomena persist in
the form of excitability. In a large domain of parameters,
an acid perturbation, made at a localized place of the gel annulus,
propagates under the form of two opposite 
travelling pulses along the inner rim.
After a short transient, each pulse takes a stationary shape
and travels without deformation along the annulus (Fig.~\ref{excit}). 
The length of the pulse tail decreases as $\alpha$ decreases until the
excitability is eventually lost. When the pulses meet, they anihilate
each other, but by adding a basic perturbation at the right place and
time , it is possible to kill one of the two pulses. Then, the
surviving pulse can rotate indefinitely along the annulus as previously
done with the Belousov-Zhabotinski reaction, the paradigm of excitable
reactions \cite{pinwheel,excyclon}. This behavior is identical to that
observed with the simple model and is also reproduced with the
extended model for the two different sets of parameters that were
checked for excitability.
   
\section{Discussion and conclusion}

It is known that, in homogeneous bistable systems, the addition of a negative
feedback that introduces a new time scale can induce oscillations.
In the same way, excitability can arise from the coupling of a
bistable subsystem that exhibits a S-shaped nulcline with the feedback
dynamics of a recovery variable.

Our simple model has clearly shown that the additional time scales
introduced by long range activation can play a similar role and are
able to induce excitability and spatio-temporal oscillations in a
system which can only be bistable in homogeneous conditions. When the
diffusion of the activator is not significantly higher than the
diffusion of the other species, the system can only exhibit spatial
bistability. But, when the ratio $d$ becomes significantly larger than
1, the system exhibits a domain of oscillations connected to the bistable
region; the diagram has a typical cross-shaped  topology \cite{note2}. 
In a narrow band within the bistability region, the `acid' state
can also become oscillatory. This is analog to the experimental observations
and to the results produced by the extended model.

Moreover, an excitability region develops along the stability limit of the FT
state. Actually, there is no discontinuity between a stable FT state
and an excitable pulse with an infinite recovery tail within the F state. 
One must be
aware that the recovery is not controlled by a slow kinetic process imbedded
in the reaction mechanism
as for the Belousov-Zhabotinskii reaction, but by the
diffusion process that rules the feed from the CSTR boundary. Thus, the
excitability properties do not only depend on intrinsic quantities, i.e. the
reaction rates and diffusion coefficients, but also on a geometrical
characteristic of the reactor, namely the depth of the gel.

Nevertheless, although the simple model predicts that differential
diffusion  can induce oscillations of the FT state in the
spatial bistability region and excitability of the F state, the
global agreement with experimental data, as well as with the extended model,
is not very good. Except for some analogies in the left upper part of
Fig.~\ref{diagvl}, the shape of the diagrams is  different. A
first reason is that the regions located on the right side of the bistable
and the `basic' domains, including the oscillation pocket, are not
physically reachable. The corresponding values of $v_0$ could not be
maintained in the CSTR with realistic residence times. The CSTR
would switch into a
thermodynamic state. By the way, with this model, it is difficult to
speak of a true FT state since the concentration profiles do not
exhibit any front. This is why we prefer to use the
terms `basic' or `acid' rather than F and FT when speaking of the
simple model. The second major reason is that OH$^-$ and the
autocatalytic species $\HP$ are related by a strong constraint
independent of the C-T reaction itself, namely the fast equilibrium of
water dissociation.  This leads to buffering processes and excludes
that $v_0$ takes values close to zero, as permitted by the simple
model. Thus this toy model, which allows for extensive numerical
calculations, is an excellent tool to demonstrate the mechanisms of
long range activation phenomena, but it should not be used for a
precise analysis of the properties of the real system. In particular,
it would be illusive to try to connect quantitatively the model parameters
with realistic quantities.

In the extended model, where the CSTR dynamics is included, a direct
comparison with experiments is possible. Spatial bistability limits
are in good agreement. There is nevertheless a point that deserves
special attention. Species $\TET$ and $\CLO$ are the natural
substrates, but since the system is fed with a basic solution, $\OH$
must be considered as another substrate in spite of its special
relation with $\HP$.  Since the diffusion coefficients of these
substrates are not equal, there is an ambiguity in the choice of the
species to be considered in the basic theory that leads to
Eq.~(\ref{valdelta}).  Fig.~\ref{numprof} clearly shows that the
relevant species is $\OH$. Its concentration drops linearly down to
very low values at the exact position of the front, whereas the other
substrates do not exhibit a linear drop all the way down and their
concentration is far from minimal at the position of the front. A
natural explanation is that, although $\OH$ diffuses fast, $\HP$
reacts much faster with $\OH$ than with the other substrates.  This
defines the smallest $\delta$ value and controls the overall
process. However, the tail of the concentration profiles of these
other subtrates plays a role in the anticipation of the
FT$\longrightarrow$F transition that is actually found in the
computations.

Like in the simple model, oscillations induced by the long range
activation are observed. If one decreases $\alpha$ within the
bistability domain, the FT state becomes oscillatory when its
stability limit is approached. The diagram is in good quantitative
agreement with experimental observations.  Since oscillations are
restricted to a narrow region of control parameters,
the unavoidable small inhomogeneities along
the annulus, make impossible to perform a fine analysis of this front
dynamics. In the model, the complex oscillations and the period
doubling sequence observed in the 1-D simulations can be understood
using the following arguments. Actually, coupled oscillators may
present the same complex dynamical behavior.  Close to the onset of
the oscillatory regime, the motion of the front exhibits simple
periodicity.  The amplitude of the oscillations $\delta_{\rm
max}-\delta_{\rm min}$ is small and the front remains located close to
the CSTR boundary. But when $\alpha$ decreases, approaching the limits of
existence of the FT state, this amplitude becomes larger and larger.
When $\delta_{\rm max}$ is of the order of $l$, the impermeable wall
acts as a reflector for the diffusive concentration fields. Then, the
front dynamics couples with itself, with delay and partial
damping. This dynamics of coupled oscillators explains the observed
dynamical complexity .

In spatial bistability, it is possible to prepare two regions of the
reactors in the two different states F and FT side by side and to
study the dynamics of the F/FT interface.  In the standard case, the
direction of propagation of this interface depends on the control
parameters and changes within the bistability domain according to the
relative stability of the two states F and FT as was experimentally
shown with the CDI reaction \cite{Bispa,BispaD}. In the C-T reaction,
this situation is radically different. Because the fast diffusion of
the autocatalytic species excites the medium ahead of the interface,
the FT state propagates always into the F state, i.e. within the whole
domain of spatial bistability.  Moreover, this behavior can persist
beyond the stability limit of state FT.  The F/FT interface still
propagates into state F, but, since the FT state is now unstable, the
system ultimately returns to the F state. This gives rise to a
recovery tail that follows the interface. If the ratio of the
diffusion of the activator to the diffusion of the main substrates is
significantly larger than 1 (here $d>3$), the propagation of the FT
state balances the return to the F state at the level of the
interface. Thus, the process goes on and a pulse can propagate
without damping in time. If $d$ is smaller,
or if the system is too far from the stability limit of the FT state,
the return to the stable state dominates and the pulse dies. This
explains the domain of excitability observed in the experiments and
accounted for by the models.

Despite the good agreement of the extended model results with experimental
data, this model still presents some serious limitations. 
A major drawback of the extended model is the use of effective
diffusion coefficients which neglects the ionic interactions. The main
effect of such interactions is more likely to give different
effective diffusion coefficients in the flow state and in the
thermodynamic state. As a result, this should change the
position of the front and the exact limits of the oscillatory and
excitability domains, but we believe that our main conclusions 
should hold. From an
experimental point on view, a severe limitation is the control of
parameter uniformity along the annular gel/CSTR boundary which makes
the front oscillations to appear in the
form of travelling waves. Finally, it has been recently found experimentally
that the kinetic mechanisms summarized by Eq.~(\ref{kin})
breaks down when the concentration
of chlorite or tetrathionate are significantly increased \cite{Itsvan}. 

Beyond these first experimental observations, one can expect to control the
front motion by a continuous change of the geometrical control parameter $l$.
Experiments along these lines are presently in
progress both numerically and experimentally.
\begin{acknowledgments}
We thank P. Borckmans, G. Dewel, K. Benyaich and A. De Wit for
discussions.
This work has been supported by the Fundaci\'on Antorchas, the C.N.R.S., and
a France-Argentina grant ECOS-SCyT.
\end{acknowledgments}


\begin{thebibliography}{99}
\bibitem{note1}
This does not preclude complex dynamics induced by the diffusive
coupling like  phase instabilities.
\bibitem{note2}
As usual, this is a first approximation of the stability limits topology.
The vicinity of the `crossing point' can exhibit more complexity like in 
\ref{diagvl}(b).
\end{thebibliography}
\end{document}